\documentclass[12pt]{article}

\usepackage{graphicx}
\usepackage{color}
\usepackage{cite}
\usepackage{amsbsy}
\usepackage{here}

\def \beq{\begin{equation}}
\def \eeq{\end{equation}}
\def\eqref#1{(\ref{#1})}
\def\bea{\begin{eqnarray}}
\def\eea{\end{eqnarray}}
\def\jpsi{J\kern-0.15em/\kern-0.15em\psi\kern0.15em}
\def\minv{M_{\rm inv}}
\def\dJp{{\rm di-}\jpsi}

\def\URLtilde{\lower0.2em\hbox{$\tilde{\phantom{a}}$}}
\def\mycomm#1{\hfill\break\strut\kern-3em{\color{red}\tt ====> #1
\color{black}}\hfill\break}

\def\Swave{$S$-wave}
\def\Pwave{$P$-wave}

\newcount\timecount
\newcount\hours \newcount\minutes  \newcount\temp \newcount\pmhours
\hours = \time
\divide\hours by 60
\temp = \hours
\multiply\temp by 60
\minutes = \time
\advance\minutes by -\temp
\def\hour{\the\hours}
\def\minute{\ifnum\minutes<10 0\the\minutes
\else\the\minutes\fi}
\def\clock{
\ifnum\hours=0 12:\minute\ AM
\else\ifnum\hours<12 \hour:\minute\ AM
\else\ifnum\hours=12 12:\minute\ PM
\else\ifnum\hours>12
\pmhours=\hours
\advance\pmhours by -12
\the\pmhours:\minute\ PM
\fi
\fi
\fi
\fi
}

\def\monthname{\relax\ifcase\month 0/\or January\or February\or
March\or April\or May\or June\or July\or August\or September\or
October\or November\or December\else\number\month/\fi}

\def\bold#1{\setbox0=\hbox{$#1$}     \kern-.025em\copy0\kern-\wd0
\kern.05em\copy0\kern-\wd0
\kern-.025em\raise.0433em\box0 }

\begin{document}
\setcounter{footnote}{1}
\vskip1.5cm
\centerline{\large \bf INTERPRETATION OF STRUCTURE}
\medskip

\centerline{\large \bf IN THE DI-$\boldsymbol{\jpsi}$ SPECTRUM}
\bigskip

\centerline{Marek Karliner$^a$\footnote{{\tt marek@tauex.tau.ac.il}}
and Jonathan L. Rosner$^b$\footnote{{\tt rosner@hep.uchicago.edu}}}
\medskip

\centerline{$^a$ {\it School of Physics and Astronomy}}
\centerline{\it Tel Aviv University, Tel Aviv 69978, Israel}
\medskip

\centerline{$^b$ {\it Enrico Fermi Institute and Department of Physics}}
\centerline{\it University of Chicago, 5640 S. Ellis Avenue, Chicago, IL
60637, USA}
\bigskip
\strut

\begin{center}
ABSTRACT
\end{center}
\begin{quote}
Structure in the $\dJp$ mass spectrum observed by the LHCb experiment around
6.9 and 7.2 GeV is interpreted in terms of $J^{PC}=0^{++}$ and $2^{++}$
resonances between a $cc$ diquark and a $\bar c \bar c$ antidiquark, using a
recently confirmed string-junction picture to calculate tetraquark masses.  The
main peak around 6.9 GeV is likely dominated by the $0^{++}(2S)$ state, a
radial excitation of the $cc$-$\bar c \bar c$ tetraquark, which we predict at
$6.871\pm 0.025$ GeV. The dip around 6.75 GeV is ascribed to the opening of the
\Swave\ di-$\chi_{c0}$ channel, while the dip around 7.2 GeV could be
correlated with the opening of the di-$\eta_c(2S)$ or $\Xi_{cc} \bar \Xi_{cc}$
channel.  The low-mass part of the $\dJp$ structure appears to require a broad
resonance consistent with a predicted $2^{++}(1S)$ state with invariant mass
around $\minv = 6400$ MeV.  Implications for $bb \bar b \bar b$ tetraquarks are
discussed.
\end{quote}
\smallskip


\section{Introduction \label{sec:intro}}

The picture of hadrons as bound states of colored quarks described the observed
mesons as $q \bar q$ and baryons as $qqq$ states, but also could accommodate
more complicated color-singlet combinations such as $qq \bar q \bar q$
(tetraquarks) or $q^4 \bar q$ (pentaquarks). Since 2003, experimental evidence
has accumulated for such combinations, but it has not been clear whether they
are genuine bound states with equal roles for all constituents, or loosely
bound ``molecules'' of two mesons or a meson and a baryon, with quarks mainly
belonging to one hadron or the other.  There is, however, fairly robust
theoretical evidence for a deeply bound genuine $bb\bar u\bar d$ tetraquark
\cite{Karliner:2017qjm,Eichten:2017ffp}.

Recently the LHCb Collaboration at CERN has presented evidence for structure in
the spectrum of a pair of $\jpsi$ mesons, $\minv(\dJp)$ \cite{Aaij:2020fnh},
interpreted as a narrow structure around 6.9 GeV and a broad structure just
above twice the $\jpsi$ mass.  A dip in $\minv(\dJp)$ around 6.75 GeV
suggests interference with nonresonant behavior in a channel with the same
$J^{PC}$.  Such behavior is difficult to regard from a molecular standpoint,
but is compatible with a picture of a compact $c c \bar c \bar c$ state.
Many theoretical interpretations of the LHCb data take this point of view
\cite{Liu:2020eha,Wang:2020ols,Jin:2020jfc,Yang:2020rih,Becchi:2020uvq,%
Lu:2020cns,Chen:2020xwe,Albuquerque:2020hio,Sonnenschein:2020nwn,%
Giron:2020wpx,Maiani:2020pur,Richard:2020hdw,Wang:2020wrp,Chao:2020dml,%
Eichmann:2020oqt,Maciula:2020wri,Faustov:2020qfm,Dong:2020nwy,%
Weng:2020jao}.

In this paper we adopt the compact tetraquark point of view (see
\cite{Karliner:2016zzc,Yang:2020atz} for lists of related predictions) and
point out a feature in the data which is characteristic of many processes.
We note that the position of the dip roughly coincides with twice the mass
of $\chi_{c0}(3415)$.  If the major resonant $\dJp$ activity is in the $J^{PC}
= 0^{++}$ channel, a pair of $\chi_{c0}(3415)$ charmonia can be produced in an
\Swave\ as soon as $\minv(\dJp)$ exceeds 6829 MeV.  Unitarity then can induce a
{\it dip} in the production channel.  (See also \cite{Albuquerque:2020hio}.)

In Section \ref{sec:sw} we recall a number of instances in which the opening
of an \Swave\ channel induces a dip in the production channel. We apply similar
methods to the \Swave\ process $\jpsi\jpsi\to\chi_{c0}\chi_{c0}$ and
$\jpsi\jpsi\to \eta_c(2S) \eta_c(2S)$ in Section \ref{sec:dips}, discuss
implications for $cc \bar c \bar c$ tetraquarks in Section \ref{sec:impc}
and for $bb \bar b \bar b$ tetraquarks in Section \ref{sec:impb},
concluding in Section \ref{sec:conc}.  An Appendix contains details of
resonance fitting.

\section{Dips and cusps in $\boldsymbol{S}$-wave production channels
\label{sec:sw}}

Dips or cusps in the cross section for a number of \Swave\ processes occur when
a new \Swave\ threshold is crossed.  Here we review several such cases.  More
details and references may be found in Ref.\ \cite{Rosner:2006vc}.

\subsection{$\boldsymbol{\pi \pi}$ $\boldsymbol{I=J=0}$ amplitude 
at $\boldsymbol{K \bar K}$ threshold}

The rapid drop in the magnitude of the $I=0$ \Swave\ \,$\pi \pi$ scattering
amplitude near a center-of-mass energy $E_{\rm cm} \simeq 1$ GeV is associated
with the rapid passage of the elastic phase shift through $180^\circ$.
(See Ref.\ \cite{Pelaez:2019eqa} for a recent parametrization.)  This behavior
is correlated with the opening of the $K \bar K$ threshold, forcing the $I=J=0$
$\,\pi \pi$ amplitude to become highly inelastic \cite{Flatte:1972rz}.  It also
reflects the effect of a narrow resonance $f_0(980)$ \cite{Zyla:2020}
coupling to both $\pi\pi$ and $K \bar K$.  For more details see
\cite{Au:1986vs,Bugg:2005nt}.  A related discussion applies to the \Swave\ $\pi
\eta$ channel near the $I=1, J=0$ $K \bar K$ threshold \cite{Flatte:1976xu}.

\subsection{Cusp in $\boldsymbol{\pi^0 \pi^0}$ spectrum
 at $\boldsymbol{\pi^+ \pi^-}$ threshold}

The $\pi^0 \pi^0$ \,\Swave\ scattering amplitude is expected to have a cusp
at $\pi^+ \pi^-$ threshold \cite{Meissner:1997fa,Meissner:1997gf}.
This behavior can be studied in the decay $K^+ \to \pi^+ \pi^0 \pi^0$, where
the contribution from the $\pi^+ \pi^+ \pi^-$ intermediate state allows one
to study the charge-exchange reaction $\pi^+ \pi^- \to \pi^0 \pi^0$ and thus
to measure the $\pi \pi$ \Swave\ scattering length difference $a_0-a_2$
\cite{Cabibbo:2005ez}.  The CERN NA48 Collaboration has performed such a
measurement, finding results \cite{Batley:2005ax} in remarkable agreement with
the prediction \cite{Cabibbo:2005ez}.  One can also study this effect in $\pi^+
\pi^-$ atoms \cite{Adeva:2005pg}.

\subsection{Hadron production by $\boldsymbol{e^+e^-}$ 
collisions around 4.26 GeV}

The value of $R \equiv \sigma(e^+ e^- \to {\rm hadrons})/\sigma(e^+ e^-
\to \mu^+ \mu^-)$ drops sharply just below threshold for production of
$D(1865)^0\bar D_1(2420)^0$ + c.c. \cite{Bai:2001ct}, which is the lowest-mass
$c \bar c$ channel accessible in an \Swave\ from a virtual photon.  If this
behavior is not coincidental, the drop in $R$ should be confined to the
$c\bar c$ final states.

\subsection{Six-pion diffractive photoproduction}

The diffractive photoproduction of $3 \pi^+\,3 \pi^-$ leads to a spectrum with a
pronounced dip near 1.9 GeV/$c^2$ \cite{Frabetti:2001ah,Frabetti:2003pw}.
This is just the threshold for production of a proton-antiproton pair in the
$^3S_1$ channel.  This dip also occurs in the $3 \pi^+\, 3 \pi^-$ spectrum
produced in radiative return in higher-energy $e^+ e^-$ collisions, i.e., in
$e^+ e^- \to \gamma \,3 \pi^+ \,3 \pi^-$, observed by the BaBar Collaboration at
SLAC \cite{Aubert:2006jq}.  The feature can be reproduced by a $1^{--}$
resonance with $M = 1.91 \pm 0.01$ GeV/$c^2$ and width $\Gamma = 37 \pm 13$ MeV
interfering destructively with a broader $1^{--}$ resonance at lower mass
\cite{Frabetti:2001ah,Frabetti:2003pw}.

\subsection{Greater generality}

The vanishing of an \Swave\ amplitude when its elastic phase shift goes through
180$^\circ$ is not confined to particle physics.  The Ramsauer-Townsend effect
represents similar behavior in atomic physics \cite{rte}.  Cusps in \Swave\
scattering cross sections occur at thresholds for {\it any} new channels
\cite{Wigner:1948,Bugg:2004rk}.  Monochromatic neutrons may be produced by
utilizing the vanishing absorption cross sections of neutrons of certain
energies on specific nuclei \cite{Barbeau:2007qh}.

\subsection{A cautionary note}

Although the rapid passage of the $I=J=0$ $\pi\pi$ phase shift through
180$^\circ$ near $K \bar K$ threshold can be ascribed to the nearby $f_0(980)$
resonance, one cannot conclude that similar behavior in other of the above
cases (or many more examined in \cite{Rosner:2006vc}) is due to nearby poles
in the scattering amplitude \cite{Bugg:2004rk}.  As in the case of
diffractive six-pion production mentioned above, unitarity alone will cause a
suppression of the input channel at the expense of the newly-open channel.
{\em The ability to fit the amplitude with a resonance does not guarantee its
existence.}

\section{Dips in $\boldsymbol{\minv(\dJp)}$ at di-charmonium thresholds
\label{sec:dips}}

\begin{figure}[t]
\begin{center}
\includegraphics[width=0.85\textwidth] {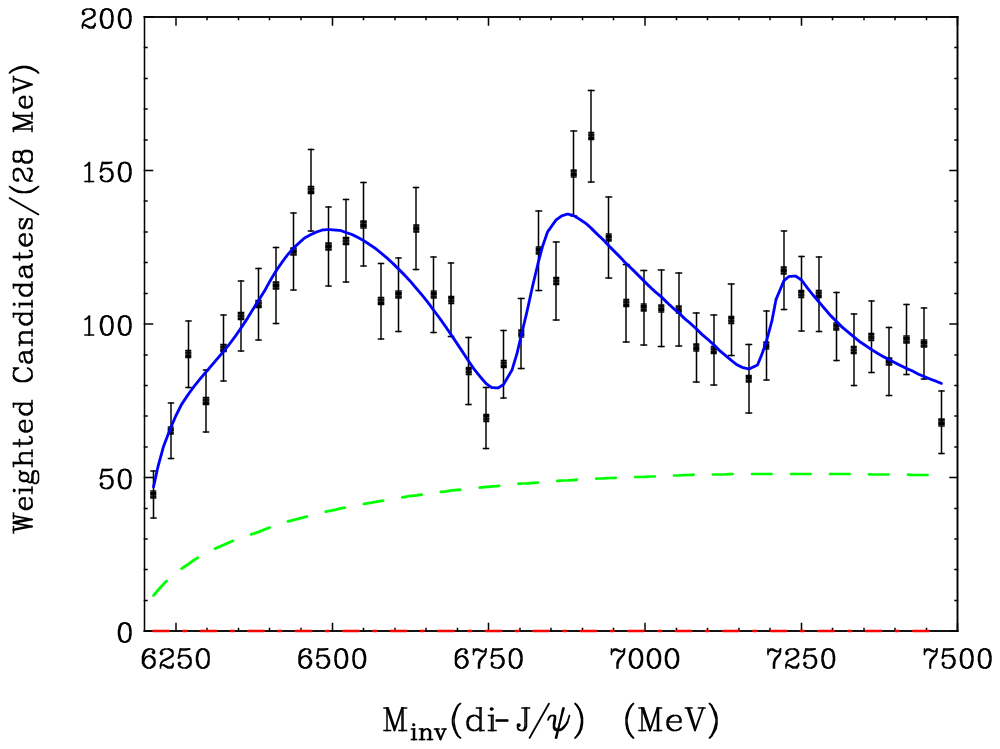}
\end{center}
\caption{Spectrum of $J/\psi$ pairs reported by the LHCb Experiment
\cite{Aaij:2020fnh}, together with our best fit to data (blue line), as given
in Table~\ref{tab:fit1} and described in the Appendix.  The green dashed line
denotes the DPS contribution, subtracted before fitting.
\label{fig:spec}}
\end{figure}

The spectrum of $\minv(\dJp)$ receives contributions from both single-parton
scattering (SPS) and double-parton scattering (DPS).  We assume, along with
\cite{Aaij:2020fnh}, that only the former process contributes to resonant
$\dJp$ structure, and subtract the latter (from \cite{Aaij:2020fnh})
before fitting the observed spectrum.  We allow a fraction $\alpha$ of the
SPS amplitude to interfere with the resonances, which are introduced by
Breit-Wigner amplitudes with arbitrary phases with respect to the nonresonant
SPS (NRSPS) amplitude, assumed real.  In Fig.\ \ref{fig:spec} we show the
spectrum together with a fit to data in the range 6.2--7.5 GeV
using the sum of three resonances with masses $M_i$, widths
$\Gamma_i$, normalizations $\eta_i$, and phases $\phi_i~(i=1,2,3)$.
Signal normalization, background normalization, and background shape are
described by parameters $C_i$ defined in the Appendix.  The results of this fit
are shown in Table \ref{tab:fit1}.

The shapes of the peaks around 6.9 and 7.2 GeV suggest destructive
interference between signal and background on the low-mass side
of both peaks.  The sudden rise following a dip is characteristic of an
S-wave amplitude, as illustrated in the previous Section.
It was associated with the opening of a nearby threshold.  In the case of
the 6.9 GeV peak, we note that $2 M(\chi_{c0}) = 6829$ MeV, so we can ascribe
the steep behavior between about 6750 and 6900 GeV as associated with opening
of the di-$\chi_{c0}$ channel.  (Ref.\ \cite{Dong:2020nwy} ascribes cusp-like
behavior to the opening of the $J/\psi~\psi(2S)$ channel at 6783 MeV.)

\begin{table}[H]
\caption{Parameters in best fit to data (see Appendix for definitions) with
  $\chi^2 = 25.855$ for 31 degrees of freedom (d.o.f.).  Masses $M_i$ and widths
$\Gamma_i$ are in MeV.  Constants $C_i$ describe signal normalization,
background normalization, and background shape, respectively.  Parameters
$\eta_i~(\eta_1 \equiv 1)$ and $\phi_i$ (in degrees) describe normalizations
and phases of $i$-th Breit-Wigner amplitudes.  
\label{tab:fit1}}
\begin{center}
\begin{tabular}{c c c c} \hline \hline
  Peak $i$   &   $i$=1  &   $i$=2  &  $i$=3  \\ \hline
\vrule width 0pt height 2.5ex
  $M_i$    & 6377.1  & 6808.6 & 7208.1 \\
  $\Gamma_i$ & 277.3 &  138.0 & 82.96 \\
   $C_i$   &  5.057  &  25.74 &  1.184 \\
 $\eta_i$  & 1.000$^a$ & 1.445 & 0.7754 \\
 $\phi_i$ & $-26.62$ & $-34.78$ & $ -4.995$ \\
 $\alpha$ & 1.000 & \multicolumn{2}{c}{Coherence factor} \\\hline \hline
\end{tabular}
\end{center}
\leftline{$^a$ Input}
\end{table}

If the di-$\chi_{c0}$ channel is in an S-wave, as implied by its sudden
onset, the S-wave behavior in the di-$\jpsi$ channel requires the two
$\jpsi$ mesons to be in a state of $J^{PC} = 0^{++}$.
An initial state of two $\jpsi$ mesons consists of two $c \bar c$ pairs,
each in a $^3S_1$ state.  A $\chi_{c0}$ is a \Pwave\ charmonium state with the
quarks' spins coupled to 1 and spin coupled with $L=1$ to give $J=0$.  The
final state with two $^3P_0$ states in a relative \Swave~can
be reached from the initial state by orbital excitation of each spin-triplet
state.  

\begin{table}
\caption{Branching fractions of $\chi_{c0}(3415)$ exceeding a percent.
\label{tab:brs}}
\begin{center}
\begin{tabular}{c c} \hline \hline
Mode & Percent \\ \hline
\vrule width 0pt height 2.5ex
$      2(\pi^+\pi^-)$     & $2.34 \pm 0.18$ \\
$\pi^+ \pi^- \pi^0 \pi^0$ &  $3.3 \pm 0.4$  \\ 
  $\pi^+ \pi^- K^+ K^-$   & $1.81 \pm 0.14$ \\
$K^+\pi^-\bar K^0\pi^0$ + c.c. & $2.49\pm0.33$ \\
      $3(\pi^+\pi^-)$     & $1.20 \pm 0.18$ \\
     $\gamma \jpsi$      & $1.40 \pm 0.05$ \\ \hline \hline
\end{tabular}
\end{center}
\end{table}

Detection of the presence of the two $\chi_{c0}$ states is challenging in view
of the small branching fractions of $\chi_{c0}$ to observable final states.
The only branching fractions of $\chi_{c0}$ that exceed a percent are
given in Table \ref{tab:brs} \cite{Zyla:2020}. With sufficient mass resolution,
one could combine the modes with all charged tracks to get an effective
branching fraction of a bit above five percent,  The total width of $\chi_{c0}$
is $10.8 \pm0.6$ MeV.  The experimental mass resolution in other LHCb analyses
(see, e.g., \cite{Aaij:2018fpa,Aaij:2019jaq}) is somewhat greater, and thus
dominates the sensitivity to a signal.  An explicit simulation would be
helpful.  If the cusp-like behavior is due to the opening of the $J/\psi~
\psi(2S)$ channel \cite{Dong:2020nwy}, that final state should be easier to
detect, consisting of two $J/\psi$ plus a $\pi^+\pi^-$ pair.

Similar behavior is apparent on the low-$\minv$ shoulder of the peak at 7.2
GeV.  The only nearby di-charmonium threshold is associated with a pair of
$\eta_c(2S)$ mesons, with $2M[\eta_c(2S)] = 7275$ MeV.  If this threshold plays
an important
role in the line shape of the peak, one should see decay products of two
$\eta_c(2S)$ mesons on the high-$\minv$ side of this peak.  This, of course, is
even more challenging than detecting a pair of $\chi_{c0}$ mesons.  (Refs.\
\cite{Sonnenschein:2020nwn,Giron:2020wpx} draw attention to the slightly lower
$\Xi_{cc} \bar \Xi_{cc}$ threshold at 7242 MeV, which we
shall discuss further at the end of Sec.\ IV.)

We initially sought evidence for a di-$\eta_c(1S)$ threshold at $2M[\eta_c(1S)]=
5968$ MeV and inserted a corresponding pole below $\dJp$ threshold into our
fitting amplitude.  The expectation was that this would contribute a needed
enhancement of the spectrum between $\minv \simeq 6.2$ and 6.6 GeV.  The
fitting program (see Appendix A) instead preferred a much higher-mass pole, as
one sees for $M_1$ in Table \ref{tab:fit1}.  Thus the predicted 1S candidate
\cite{Karliner:2016zzc} with mass $M[T(c c \bar c \bar c)] = 6191.5 \pm 25$ MeV
for the lightest all-charm tetraquark remains to be observed.  The value of
$M_1$ in Table \ref{tab:fit1} is more consistent with that of a $2^{++}(1S)$ 
state predicted in Ref.\ \cite{Faustov:2020qfm} to lie at 6367 MeV. 
These authors predict a mass of 6190 MeV for the 1S state, consistent with ours.

Although we do not predict a tetraquark resonance near di-$\eta_c(1S)$
threshold, it would be worth examining channels that couple to a pair of
$\eta_c(1S)$ to see if they exhibit cusps in S-wave amplitudes near $\minv =
5968$ MeV.  Examples of such channels include $D \bar D$ and $D^* \bar D^*$
\cite{Becchi:2020uvq,Maiani:2020pur}.

\section{Implications for $cc \bar c \bar  c$ tetraquarks \label{sec:impc}}

In Ref.\ \cite{Karliner:2016zzc}, using a diquark-antidiquark picture,
we predicted the ground state $T(c c \bar c \bar c)$ mass to be $6191.5
\pm 25$ MeV.  This error is taken to be twice that obtained when fitting
non-exotic mesons and baryons in the string-junction picture (see also
\cite{Sonnenschein:2020nwn}), recently confirmed by the successful prediction
of the mass of a $T(cs \bar u \bar d)$ tetraquark \cite{Karliner:2020vsi} and
which we are assuming here \cite{Karliner:2016zzc}.
This would be the $0^{++}(1S)$ state of the spin-1 color
antitriplet diquark and the spin-1 color triplet antidiquark.  The
ingredients of the prediction included a term $2S = 2(165.1)$ MeV for two
 QCD string junctions, $2(M_{cc}) = 2(3204.1)$ MeV for the masses of two
diquarks, an interpolated binding energy of the $cc$ diquark with the
$\bar c \bar c$ antidiquark of --388.3 MeV, and a hyperfine term of
  --158.5 MeV.  The predicted mass is just below $2M(\jpsi) = 6194$ MeV
but above $2M(\eta_c(1S) = 5968$ MeV, so strong decay to a pair of
$\eta_c(1S)$ is favored.  Here and subsequently we use the latest
Particle Data Group masses \cite{Zyla:2020}.

The above discussion is based on S-wave $cc$ diquarks in a color $3^*$ state,
with spin 1.  There should also be states involving color 6 diquarks, with
spin zero.  There should be an additional spinless tetraquark made of a 6
in an S-wave state with a $6^*$.  Estimates, for example in Ref.\
\cite{Wang:2020wrp}, of its mass are not far from that of the 1S $3^* \times 3$
state, and the two may mix with one another.

The above estimate concerns the ground state $0^{++}$ mass.  One estimates the
ground state $2^{++}$ mass by noting that the hyperfine terms for a pair of
spin-1 particles in states of $J = 0,1,2$ are in the ratio $(1/2)[J(J+1) - 4] =
-2, -1, 1$, so the hyperfine term for the lowest $2^{++}$ state is 79.3 MeV and
the mass of the $2^{++}(1S)$ state is $6429 \pm 25$
MeV, 238 MeV above the $0^{++}(1S)$ and well above $2M(\jpsi)$ threshold.
This is the state predicted in Ref.\ \cite{Faustov:2020qfm} to lie at 6367
MeV.  It could be some or all of the low-$\minv$ $\dJp$ peak with $M_1 =
6377$ MeV,in Table \ref{tab:fit1}, allowing the $0^{++}$ state to lie at our
predicted value of $6191.5 \pm 25$ MeV.  A spin-parity analysis should be able
to distinguish between $0^{++}$ and $2^{++}$ components of the amplitude.

The 1S--2S splittings of the charmonium and bottomonium systems are
almost the same.  The spin-weighted average (1S, 2S) masses are (3068.65,
3673.95) MeV for charmonium and (9444.9, 10017.2) MeV for bottomonium,
so $M(2S) - M(1S) = (605.3, 572.3)$ MeV for ($c \bar c,~b \bar b$). 
They would be equal for a logarithmic interquark potential, providing a
convenient interpolation between short-distance and long-distance QCD for
these systems \cite{Quigg:1977dd}.  The $cc$ diquark mass is intermediate
between $m_c$ and $m_b$: using the values from \cite{Karliner:2016zzc},
\beq \label{eqn:m}
m_c = 1655.6~{\rm MeV}~,~~m_b = 4988.6~{\rm MeV}~,~~m_{cc} = 3204.1~{\rm MeV}~,
\eeq
a power-law interpolation between $m_c$ and $m_b$ of the form $M(2S) - M(1S)
= a m^p$ with $a = 882.22 m^{-0.050826}$  gives the 1S--2S splitting for a $cc$
diquark and a $\bar c \bar c$ antidiquark to be 585.3 MeV.

The hyperfine splittings $M(^3S_1) - M(^1S_0)$ are in the ratio
$\Delta M(2S)/\Delta M(1S) = 0.430\pm0.005$ for $c \bar c$ 
and $0.390\pm0.066$ for $b \bar b$.
Interpolating these central values in terms of a power law in masses
(\ref{eqn:m}) we find $\Delta M(2S)/\Delta M(1S) = 0.4053$
for the bound states of the $cc$ diquark and the $\bar c
\bar c$ antidiquark.  This means that for the 2S system, we replace the
 1S hyperfine term of --158.5 MeV by --64.2 MeV, a change of 94.3 MeV.  The
mass of the $0^{++}(2S)$ state is then $6192 + 585 + 94 = 6871$ MeV,
close to the peak claimed by LHCb. (See also \cite{Lu:2020cns,Giron:2020wpx}.)
The $2^{++}(2S)$ state is then (0.4053)(237.8) = 96 MeV higher, at 6967 MeV.
This state could also be contributing to the LHCb signal.

We have not discussed $1^{++}$ states of $cc$ diquark and $\bar c \bar c$
antidiquark decaying to a pair of $\jpsi$ in an \Swave.  Two identical
spin-1 bosons in an \Swave\ are forbidden by Bose statistics to have total
angular momentum $J=1$.  Various predictions have been made for the mass of
a $1^{+-}$ state, which, however, does not couple to a pair of $J/\psi$.
The \Swave\ threshold amplitude amplitude  for $\chi_{c0} \chi_{c0}$
production, starting at $2 M(\chi_{c0}) = 6829$ MeV, thus interferes
primarily with the $0^{++}(2S)$ $2\jpsi$ resonant amplitude.

The peak around 7200 MeV is in approximately the right place for a 3S state
of $(cc)_{3^*} (\bar c \bar c)_3$.
The flavor threshold for charmonium lies just above 
the 2S level, while that for bottomonium lies just below the 4S level.
As a system with reduced mass intermediate between that of charmonium and
that of bottomonium, the $\dJp$ system can be expected to have a flavor
threshold around the 3S level (see Fig.\ 1 of\cite{Quigg:1977xd}).  This
estimate is based on the observation \cite{Rosner:1996xz,Rosner:2014nka} that
flavor threshold in a quarkonium system always occurs at a universal length of
the QCD string connecting the two heavy constituents.
Indeed, the first open-flavor state in which a QCD string connecting
$(cc)_{3^*}$ with $(\bar c \bar c)_3$ breaks is that in which a light $q 
\bar q$ pair is produced, giving $\Xi_{cc} \bar \Xi_{cc}$ with threshold 
7242 MeV \cite{Sonnenschein:2020nwn,Giron:2020wpx}.

\section{Implications for $bb \bar b \bar b$ tetraquarks
\label{sec:impb}}

Some attention to the question of fully heavy tetraquarks was drawn by an
unpublished report by the CMS Collaboration at CERN \cite{Durgut:2017}
of an exotic structure in the four-lepton channel at $18.4 \pm 0.1 \pm 
0.2$ GeV, an excess with a global significance of 3.6 $\sigma$. 
CMS reported $38 \pm 7$ events of $\Upsilon(1S)$ pairs produced
with an integrated luminosity of 20.7 fb$^{-1}$ at $\sqrt{s} = 8$ TeV, each
decaying to $\mu$ pairs \cite{Khachatryan:2016ydm}.  
There is no published
confirmation of the structure \cite{Aad:2015oqa,Khachatryan:2017mnf}, but in
view of the $\dJp$ structure it is worth updating and extending the predictions
of Ref.\ \cite{Karliner:2016zzc} for $bb \bar b \bar b$ tetraquarks.

In Ref.\ \cite{Karliner:2016zzc} we predicted the ground state $T(bb \bar b
\bar b)$ mass to be $18826 \pm 25$ MeV, just above $2M[\eta_b(1S)] =  18797$
MeV, so its main decay will likely be to two $\eta_b$-s.  It would be the
$0^{++}$ state of a color antitriplet spin-1 $bb$ diquark and the corresponding
antidiquark.  One predicts
\bea
M[T(bb\bar b \bar b)(0^{++})] & = & 2S + 2 M(bb,3^*) + B_{(bb)(\bar b \bar b)}
 + \Delta M_{HF} \nonumber \\ 
& = & [2(165.1) + 2(9718.9) -855.7 - 86.7]~{\rm MeV} = 18{,}825.6~{\rm MeV}~,
\eea
where $S$ is the contribution of a QCD string junction, $B_{(bb)(\bar b \bar
b)}$ is the binding energy between the $bb$ diquark and the $\bar b \bar b$
antidiquark, and $\Delta M_{HF}$ is the hyperfine interaction between the
diquark and the antidiquark.  An error of $\pm 25$ MeV was assigned to this
prediction, which we will assume applies to the other predictions in this
Section.

The hyperfine term for the $2^{++}$ state is $(-1/2)(-86.7) = 43.4$ MeV, so the
$2^{++}$ (1S) state is 130.1 MeV higher than the $0^{++}$ (1S) state, or
18955.7 MeV.  This lies above $2M(\Upsilon(1S) = 2(9460.3) = 18920.6$ MeV
so it can decay to a pair of $\Upsilon(1S)$.

In order to estimate the 1S--2S splitting for $T(bb \bar b \bar b)$, we 
use the power-law dependence of the previous Section, $\Delta M = 882.22
m^{-0.050826}$ (units in MeV) with $m = 9718.9$ MeV, to predict $M(2S) -
M(1S) = 553.2$ MeV.  To estimate the 2S hyperfine splitting we extrapolate
the ratio $\Delta M_{HF}(2S)/\Delta M_{HF}(1S) = 0.83232 m^{-0.089428}$
to obtain $\Delta M_{HF}(2S)/\Delta M_{HF}(1S) = 0.3671$.  The hyperfine terms
for $(0^{++},2^{++})(2S)$ are then $(-31.8,15.9)$ MeV, resulting in the
predictions $M(0^{++},2^{++})(2S) = (19433.6,19481.4)$ MeV.

The radially excited $0^{++}(2S)$ $bb$-$\bar b \bar b$ tetraquark at
$19.434 \pm 0.025$ GeV is the bottom analogue of the $0^{++}(2S)$ excited 
$cc$-$\bar c \bar c$ tetraquark at $6.871\pm 0.025$ GeV, proposed here
as the main component of the peak near 6.9 GeV reported by
LHCb \cite{Aaij:2020fnh}.

The predicted $0^{++}(2S)$ mass is large enough to imply a substantial partial
width into a pair of $\Upsilon(1S)$.  It lies below the $\chi_{b0}\chi_{b0}$
threshold, which is $2(9859.44) = 19718.9$ MeV, so its interference with the
$0^{++}$ state will depend on the width of that state and should exhibit
a different pattern from the $T(cc \bar c \bar c)$ case, where the $\chi_{c0}
\chi_{c0}$ threshold roughly coincides with the $0^{++}(2S)$ resonance mass. 
We should also keep in mind the $\Xi_{bb} \bar \Xi_{bb}$ threshold at
$2(10162 \pm 12) = 20324 \pm 25$ MeV, where we have used the prediction
\cite{Karliner:2014gca} $M(\Xi_{bb}) = 10162 \pm 12$ MeV, in analogy with the
$\Xi_{cc} \bar \Xi_{cc}$ threshold mentioned earlier.

\section{Conclusions \label{sec:conc}}

We have interpreted the structure in the $\dJp$ mass spectrum observed by
LHCb in terms of a diquark-antidiquark picture
\cite{Karliner:2016zzc}, with the predicted masses in Table \ref{tab:cpreds}.
\begin{table}
\caption{Predicted masses of lowest-lying bound states of a color-antitriplet
spin-1 $cc$ diquark and a color-triplet spin-1 $\bar c \bar c$ antidiquark.
The $\chi_{c0} \chi_{c0}$ threshold is 6829 MeV.
\label{tab:cpreds}}
\begin{center}
\begin{tabular}{c c c} \hline\hline
 & $M(1S)$ (MeV)  & $M(2S)$ (MeV) \\ \hline
\vrule width 0pt height 2.5ex
$J^{PC} = 0^{++}$ & $6192 \pm 25$ & $6871 \pm 25$ \\
$J^{PC} = 2^{++}$ & $6429 \pm 25$ & $6967 \pm 25$ \\ \hline \hline
\end{tabular}
\end{center}
\end{table}
The irregular structure is seen to be due to the rapidly opening $\chi_{c0}
\chi_{c0}$ \Swave\ channel at 6829 MeV, interfering primarily with the $0^{++}$
$2S$ state.  Anther possibility is the opening of the $J/\psi$ $\psi(2S)$
threshold at 6783 MeV \cite{Dong:2020nwy}, which should show up in the $2\mu^+
2\mu^- \pi^+ \pi^-$ final state.  We have also updated and extended our
prediction \cite{Karliner:2016zzc} for the tetraquark $T(bb \bar b \bar b)$,
with the results shown in Table \ref{tab:bpreds}.
\begin{table}
\caption{Predicted masses of lowest-lying bound states of a color-antitriplet
spin-1 $bb$ diquark and a color-triplet spin-1 $\bar b \bar b$ antidiquark.
The $\chi_{b0} \chi_{b0}$ threshold is 19719 MeV.
\label{tab:bpreds}}
\begin{center}
\begin{tabular}{c c c} \hline\hline
 & $M(1S)$ (MeV)  & $M(2S)$ (MeV) \\ \hline
\vrule width 0pt height 2.5ex
$J^{PC} = 0^{++}$ & $18826 \pm 25$ & $19434 \pm 25$ \\
$J^{PC} = 2^{++}$ & $18956 \pm 25$ & $19481 \pm 25$ \\ \hline \hline
\end{tabular}
\end{center}
\end{table}
The relative position of the $2\chi_{b0}$ threshold with respect to the
predicted $0^{++}(2S)$ state is different from that in the charm case,
implying a structure in invariant mass of different shape.

\section*{Acknowledgements}
\vskip-0.2cm
We thank Liupan An for discussion of the LHCb data following the LHC seminar on
June 16 and for directing us to the supplemental files with numerical values
and errors in Ref.~\cite{Aaij:2020fnh}.  We also thank Richard Lebed, Mikhail
Mikhasenko, Tomasz Skwarnicki, and Changzheng Yuan for useful discussions.
The research of M.K. was supported in part by NSFC-ISF grant No.\ 3423/19.
\vskip-1.0cm\strut

\section*{Appendix:  Details of data fitting}
\vskip-0.2cm
We assume the $\dJp$ spectrum is due to a smooth background with proper
threshold behavior:
\beq
B(\minv) = - C_2 q \exp[(2M(J/\psi) - \minv) ({\rm GeV}) C_3]~,~~
q \equiv (\minv^2/4 - [M(J/\psi)]^2)^{1/2}~,
\label{bkg:eq}
\eeq
of which an amplitude fraction $\alpha$ is
added coherently to the sum of three Breit-Wigner resonances each of the form
\bea
A_i & = & N_i/D_i~,~~N_i = C_1 e^{i\phi_i}\eta_i \minv \Gamma_i~, \nonumber \\
D_i & = & M_i^2 - \minv^2 - i \minv \Gamma_i~,~~(i = 1,2,3)~,
\eea
where $M_i$ and $\Gamma_i$ are the mass and width of the $i$th resonance.  The
best fit is obtained for $\alpha = 1$, consistent with the assumption in
Model II of Ref.\ \cite{Aaij:2020fnh}.  We set $\eta_1 \equiv 1$ and absorb
normalization of resonance 1 into the constant $C_1$.  The constants $C_2$ and
$C_3$ parametrize background normalization and shape, respectively.  The
observed number of events per 28 MeV bin is then
\beq
N(\minv) = |T(\minv)|^2~,~~T \equiv B + \sum_1^3 A_i~.
\eeq
The numerical data $N \pm dN$ are those in Fig.\ 3(a) of Ref.\
\cite{Aaij:2020fnh}, restricted to the range $6200 \le \minv \le 7488$ MeV
(our choice of upper bound; the data are quoted up to 8000 MeV).  We minimize
$\chi^2 \equiv \sum_j \{ [N_j({\rm fit}) - N_j({\rm data})] / dN_j \}^2$, the
sum over 46 28-MeV-wide bins centered on from 6214 to 7474 MeV.

Some parameters are not well determined by the $\chi^2$  criterion, and must
be regarded as only representative values.  To illustrate this, we present
in Table \ref{tab:fit2} the best fits for $\alpha = 0.7156$
(a local $\chi^2$ minimum with $\chi^2 = 25.86787$ for 32 d.o.f.) and $\alpha =
0$ (giving the largest global $\chi^2$ minimum, $\chi^2 = 26.19538$, for any
fixed value of $\alpha$ between 0 and 1.
\vskip-0.6cm\strut
\begin{table}[H]
\caption{Parameters in alternative fits to data with $\alpha = 0.7156$ (left)
and 0 (right).  Notation as in Table \ref{tab:fit1}.
\label{tab:fit2}}
\begin{center}
\begin{tabular}{c c c c c c c} \hline \hline
 &\multicolumn{3}{c}{$\alpha=0.7156$} & \multicolumn{3}{c}{$\alpha=0$} \\
  Peak $i$ & $i$=1 & $i$=2 & $i$=3 & $i$=1 & $i$=2 & $i$=3  \\ \hline
\vrule width 0pt height 2.5ex
  $M_i$ & 6377.9 & 6898.4 & 7208.4  & 6500.0  & 6811.3 & 7211.8 \\
  $\Gamma_i$ & 286.3 & 138.8 & 81.85 & 729.2 & 155.9 & 92.21 \\
   $C_i$   &  24.65  &  25.66 &  1.245 & 19.25 & 16.62 & 7.019 \\
 $\eta_i$  & 1.000$^a$ & 0.5227 & 0.1902 & 1.000$^a$ & 0.4666 & 0.2097 \\
 $\phi_i$ & -74.37  & -95.02 &  -51.39 & 37.61 & -46.87 & 9.263\\ \hline \hline
\end{tabular} 
\end{center}
\vskip-0.4cm
\leftline{\strut\kern6.4em $^a$ Input}
\end{table}

\end{document}